\documentstyle[floats,preprint,pra,epsfig,aps]{revtex}
\begin{document}
\draft
\renewcommand{\textfraction}{0.0}
\renewcommand{\dblfloatpagefraction}{0.8}
\renewcommand{\topfraction}{1.0}
\renewcommand{\bottomfraction}{1.0}
\renewcommand{\floatpagefraction}{0.8}
\def\vep{\varepsilon}
\def\ep{\epsilon}
\def\al{\alpha}
\def\be{\beta}
\def\ga{\gamma}
\def\la{\lambda}
\def\th{\theta}
\def\de{\delta}
\def\si{\sigma}
\def\ti{\tilde}
\def\et{\eta}
\def\pa{\partial}
\def\om{\omega}
\def\fr{\frac}
\def\be{\begin{equation}}
\def\ee{\end{equation}}
\def\bea{\begin{eqnarray}}
\def\eea{\end{eqnarray}}
\title{Quantum Depletion
       of a Soliton Condensate}
\author{Guoxiang Huang$^{1,2,3}$, L. Deng$^2$, Jiaren Yan$^{4}$, and Bambi Hu$^{3,5}$}
\address
{$^{1}$Department of Physics,
      East China Normal University, Shanghai 200062, China\\
$^{2}$Electron \& Optical Physics Division, NIST, Gaithersburg, MD 20899\\
$^{3}$Department of Physics and Centre for Nonlinear Studies, Hong
      Kong Baptist University, Hong Kong, China\\
$^{4}$Department of Physics, Hunan Normal University, Changsha,
      Hunan 410083, China\\
$^{5}$Department of Physics, University of Houston, Houston, TX 77204}

\maketitle
\begin{abstract}
We present rigorous results on the diagonalization of Bogoliubov
Hamiltonian for a soliton condensate. Using the complete and
orthonomalized set of eigenfunction for the Bogoliubov de Gennes
equations, we calculate exactly the quantum depletion of the
condensate and show that two degenerate zero-modes, which
originate from a $U(1)$ guage- and a translational-symmetry
breaking of the system, induce the quantum diffusion and
transverse instability of the soliton condensate.

\vspace{2mm}

\end{abstract}

The study of matter waves and elementary excitations has received
much attention because of the remarkable experimental realization
of Bose-Einstein condensation in trapped, weakly interacting
atomic gases\cite{pet,pit,den}.  In the past few years, several
paths have been explored in studying elementary excitations in
Bose-Einstein condensates (BECs).  The most followed method is to
use the Gross-Pitaevskii (GP) equation which is suitable to
describe a zero-temperature BEC. One of the deficiency of this
approach, in addition to the constraint of zero temperature, is
the neglect of the quantum fluctuations of condensate, which is an
important aspect of elementary excitation\cite{stri}. The
eigen-modes of the elementary excitations thus obtained so
far\cite{pet,pit} are not complete and their orthonormalities have
never been proved.

A different approach on elementary excitations is to use
Bogoliubov theory\cite{bog} that was originally formulated for
homogeneous Bose systems but is also valid for {\it inhomogeneous}
ones. In this approach one makes a canonical transformation for
boson operators to diagonalize the quantum Hamiltonian of system
and hence the quantum fluctuations of condensate are taken into
account\cite{lew}. An added advantage of the Bogoliubov theory is
that it can be easily generalized to the case of finite
temperature (with, e. g., the Hatree-Fock and Popov
approximations)\cite{pet,pit}, therefore can be compared directly
with experiments carried out at non-zero temperature environment.

In this Letter we apply the Bogoliubov theory to the investigation
of the elementary excitations generated from a quasi
one-dimensional(1D) soliton condensate. It is known that a quasi
1D BEC can form in the presence of a transverse confinement, which
introduces a cutoff for long wavelength fluctuations in the
transverse directions and hence an off-diagonal-long-range order
can establish, as have been reported in the case of low-dimensions
BEC systems\cite{gor} and in the case of matter-wave bright
solitons in quasi-1D systems\cite{kha,str} and the corresponding
flourished theoretical activities\cite{car,khaw,kan,kav}. The main
contribution of our present work is the rigorous diagonalization
of quantum Bogoliubov Hamiltonian and the exact calculation of
quantum depletion of the soliton condensate and related
conclusions on soliton stability.


We consider a quasi 1D attractive, weakly interacting Bose gas
condensed in a trap with tight transverse confinement and
negligible trapping potential in the axial direction. The grand
canonical Hamiltonian of the system can be written as the
following dimensionless from
\be \label{gham}
\hat{K}=\hat{H}-\mu \hat{N}= \int d z \hat{\Psi}^{\dag}(z,t)\left[
-\fr{\pa^2}{\pa z^2}-\mu-\hat{\Psi}^{\dag}(z,t)
\hat{\Psi}(z,t)\right] \hat{\Psi}(z,t),
\ee
where $\hat{\Psi}(z,t)$ is the field operator satisfying
$[\hat{\Psi}(z,t),\hat{\Psi}^{\dag}(z^{\prime},t)]=\delta
(z-z^{\prime})$ (with other commutators being zero). The chemical
potential $\mu$ assures the conservation of the average particle
number $\hat{N}=\int dz \hat{\Psi}^{\dag}(z,t)\hat{\Psi}(z,t)$. In
the expression (\ref{gham}) the coordinate, time, field operator,
and energy (including chemical potential) are scaled by
$l=\hbar^2/(m |U_0|)$, $2ml^2/\hbar$, $l^{-1/2}$, and
$\hbar^2/(2ml^2)$, respectively. $U_0(<0)$ is the interatomic
interaction constant and $m$ is the atomic mass. The model
(\ref{gham}) has been used in Ref.\cite{kan} to study the quantum
phase transition from a homogeneous ground state to a
bright-soliton ground state when the strength of the attractive
interaction is increased.


Letting $\hat{\Psi}(z,t)=\psi_g(z)+\hat{\psi}(z,t)$ and assuming
$||\hat{\psi}||<<\psi_g$ we obtain the Bogoliubov form (quadratic
for $\hat{\psi}$ and $\hat{\psi}^{\dag}$) of the grand canonical
Hamiltonian  $\hat{K}=K_g+\hat{K}_2$ with $K_g=\int dz \psi_g
(-\pa^2/\pa z^2-\mu-\psi_g^2 )\psi_g$ and $\hat{K}_2=\int dz
\left[ \hat{\psi^{\dag}} (-\pa^2/\pa
z^2-\mu-4\psi_g^2)\hat{\psi}-\psi_g^2
(\hat{\psi}\hat{\psi}+\hat{\psi}^{\dag}\hat{\psi}^{\dag})\right]$,
where the ground-state wave function satisfies the equation $(
-\pa^2/\pa z^2-\mu-2\psi_g^2)\psi_g=0$. For small $|U_0|$ the
ground state is a homogeneous condensate and hence the Hamiltonian
can be easily diagonalized and the elementary excitation exhibits
a gapless phonon spectrum.  For large $|U_0|$, however, this
homogeneous condensate is dynamically unstable\cite{kan} and the
system undergoes a quantum phase transition into the soliton
ground state
\be \label{soliton}
\psi_g(z)=(N_0/2){\rm sech}[(N_0/2) (z-z_0)]
\ee
with chemical potential $\mu=-N_0^2/4$, where $z_0$ is an
arbitrary constant, indicating that the quantum phase transition
results in a translational symmetry breaking of the system.

Our first objective is to search for a rigorous diagonalization of
the Hamiltonian of the soliton condensate (\ref{soliton}). Letting
$(N_0/2) (z-z_0)\rightarrow z$ and
$\hat{\psi}=(N_0/2)^{1/2}\hat{\phi}$  we get
\be \label{solH2}
\hat{K}=K_g+(N_0^2/4)\int dz \left[ \hat{\phi}^{\dag}
\hat{L}\hat{\phi}-{\rm sech}^2
z(\hat{\phi}\hat{\phi}+\hat{\phi}^{\dag}\hat{\phi}^{\dag})\right],
\ee
where $\hat{L}=-\pa^2/\pa z^2-4{\rm sech}^2 z+1$ and $\hat{\phi}$
satisfies
$[\hat{\phi}(z,t),\hat{\phi}^{\dag}(z^{\prime},t)]=\delta
(z-z^{\prime})$ (with other commutators being zero). To
diagonalize $\hat{K}$ we make the canonical transformation
\be
\label{bogt} \hat{\phi}=\sum_j \left[
u_j(z)\hat{c}_j+v_j^*(z)\hat{c}_j^{\dag}\right] +\sum_{k\neq
0}\left[ u_k(z)\hat{c}_k+v_k^*(z)\hat{c}_k^{\dag}\right],
\ee
where $\hat{c}_j$ and $\hat{c}_k$ are usual boson operators. The
first and second terms on the right-hand-side (RHS) of
(\ref{bogt}) are respectively the contributions from the discrete
($j$) and continuum ($k$) spectra of the excitations generated
from the soliton condensate.  The key for diagonalizing $\hat{K}$
is to find a complete set of the eigen-functions $\{u_q(z),v_q(z);
q=k,j \}$ with which the expansion of $\hat{\phi}$ can be made.
The discrete modes are in fact the eigen-modes with a zero
eigenvalue, as shown below.

Substituting (\ref{bogt}) into (\ref{solH2}) and assuming that
$u_q$ and $v_q$ fulfill the following eigenequations:
\be
\label{eigeneq}\hat{L}_2\hat{L}_1\phi_q(z)=E_q^2\,
\phi_q(z),\hspace{0.7cm}
\hat{L}_1 \hat{L}_2 \psi_q(z)=E_q^2\, \psi_q(z),
\ee
where $\psi_q (z)=u_q (z)+v_q (z)$, $\phi_q (z)=u_q (z)-v_q (z)$
and $\hat{L}_j=d^2/dz^2+(2\delta_{j1}+6\delta_{j2}){\rm sech}^2
z-1$ ($j=1,2$), by a detailed calculation we obtain a diagonalized
$\hat{K}=K_g+(N_0^2/4)\hat{K}_{20}$, with
\be \label{diaham2}
\hat{K}_{20}=\fr{2}{3}-\sum_{k\neq 0}\int dz E_k |v_k(z)|^2
-\hat{P}_1^2+\hat{Q}_2^2 +\sum_{k\neq 0} E_k
\hat{c}_k^{\dag}\hat{c}_k,
\ee
where we have introduced the operators\cite{lew}
$\hat{P}_j=(\hat{c}_j+\hat{c}_j^{\dag})/\sqrt{2}$ and
$\hat{Q}_j=i(\hat{c}_j-\hat{c}_j^{\dag})/\sqrt{2}$. Equation
(\ref{eigeneq}) is equivalent to the Bogoliubov-de Gennes (BdG)
eigen-value problem:
\bea
& & \label{bdg1} \hat{L}u_q(z)-2{\rm sech}^2z\, v_q(z)=E_{q}
u_q(z),\\
& & \label{bdg2} \hat{L}v_q(z)-2{\rm sech}^2z\,
u_q(z)=-E_{q}v_q(z).
\eea
The solutions of the Eq. (\ref{eigeneq}) has been known in {\it
classical} soliton perturbation theory\cite{kau,yan}. Thus we can
use the result obtained in Refs. \cite{kau,yan} to get the
following solutions of the BdG equations (\ref{bdg1}) and
(\ref{bdg2}):
\bea
& & \label{eigfun1}
u_k(z)=-\fr{1}{\sqrt{2\pi}(k^2+1)} \left[k^2+2ik \tanh
z-2\tanh^2z\right] \exp (ikz),\\
& & \label{eigfun2}
v_k(z)=-\fr{1}{\sqrt{2\pi}(k^2+1)}\, {\rm sech}^2 z\exp (ikz)
\eea
with
\be \label{spectrum}
E_k=k^2+1
\ee
for $k\neq 0$ (continuous spectrum). Equations (\ref{bdg1}) and
(\ref{bdg2}) admit also the following zero-mode solutions
($j=1,2$):
\bea
& & \label{zeromode1} u_j(z)=\delta_{j1}\,{\rm sech}z\,(2-z\tanh
z)/2+\delta_{j2}\, {\rm sech}z\,(z+\tanh z)/2,\\
& & \label{zeromode2} v_j(z)=-\delta_{j1}\,(z/2)\,{\rm
sech}z\,\tanh z+\delta_{j2}\,{\rm sech}z\,(\tanh z-z)/2.
\eea
These zero-modes belong to the discrete spectrum of the system
with a zero eigenvalue, which is two-fold degenerate. Physically,
such degenerate zero-modes originate from a $U(1)$ guage- as well
as a translational-symmetry breaking of the system and thus open a
gap in the excitation spectrum (see Eq. (\ref{spectrum})\,). This
is very different from the result obtained for homogeneous
condensates. It can be shown that the eigenfunction set obtained
above form a complete function set and they are also
orthogonal\cite{note}.

With the above results we find that $K_g=2N_0^3/3$ and
$\sum_{k\neq 0}E_k\int dz |v(z,k)|^2=(\pi-1)/(6\pi)$ and obtain
diagonalized  (dimensional) Bogoliubov quantum Hamiltonian
\be \label{Hamiltonian}
\hat{H}=-\fr{1}{24} \fr{m |U_0|^2}{\hbar^2} N_0^3+\fr{1}{8}\fr{m
|U_0|^2}{\hbar^2} N_0^2 \left[
\fr{2}{3}+\fr{\pi-1}{6\pi}-\hat{P}_1^2+\hat{Q}_2^2+\sum_{k\neq 0}
E_k \hat{c}_k^{\dag}\hat{c}_k\right].
\ee
Note that each term on the RHS of (\ref{Hamiltonian}) has clear
physical meaning. The first term describes the ground state energy
of the system, which for large $N_0$ agrees with the result of
Bethe ansatz solution\cite{mcg}. The first term in the square
bracket comes from the zero-modes whereas the second term is the
contribution of the depletion of the condensate. The terms
$-\hat{P}_1^2+\hat{Q}_2^2$ are originated from the quantum
fluctuations of the condensate and the last term represents the
contribution from phonons. Because the continuous spectrum of the
excitations $E_k$ for the soliton condensate opens a gap, thus the
phonons are {\it massive} due the breaking of translational
symmetry of the system. This is very different from the case of
homogeneous condensates where a phonon is non-massive.

Our next objective is to investigate the quantum dynamics of the
soliton condensate. For the zero-modes we obtain the equations of
motion
\bea
& & d\hat{P}_1/d\tau=0,\hspace{1cm}
d\hat{Q}_1/d\tau=-2\hat{P}_1,\\
& & d\hat{P}_2/d\tau=-2\hat{Q}_2, \hspace{1cm}
d\hat{Q}_2/d\tau=0,
\eea
where $\tau=(N_0^2/4) t$. The Hermitian operators $\hat{P}_j$ and
$\hat{Q}_j$  satisfy the commutation relation
$[\hat{Q}_j,\hat{P}_{j^{\prime}}]=i\delta_{jj^{\prime}}$, and they
are associated with the collective motion of the solition
condensate\cite{lew}. The exact solutions of the above equations
read $\hat{P}_1(\tau)=\hat{P}_1(0)$,
$\hat{Q}_1(\tau)=\hat{Q}_1(0)-2\hat{P}_1(0)\tau$,
$\hat{P}_2(\tau)=\hat{P}_2(0)-2\hat{Q}_2(0)\tau$, and
$\hat{Q}_2(\tau)=\hat{Q}_2(0)$.  Using the zero-mode solutions and
the relations between $\hat{c}_j$, $\hat{c}_j^{\dag}$ and
$\hat{P}_j$, $\hat{Q}_j$  we obtain
\be
\hat{\Psi}(z,t)\approx \hat{A}(z,\tau)\, {\rm sech}\left[
z-\fr{1}{\sqrt{N_0}} \hat{P}_2(\tau)\right]  \exp \left\{
-\fr{i}{\sqrt{N_0}} \left[ \hat{Q}_1(\tau)+z
\hat{Q}_2(\tau)\right]\right\},
\label{fieldoperator}
\ee
with $\hat{A}(z,\tau)\approx (N_0/2) \left[1+ (1-z \tanh
z)\hat{P}_1 (\tau)/\sqrt{N_0}\right]$. Equation
(\ref{fieldoperator}) clearly shows quantum fluctuations in the
amplitude, position and phase of the soliton condensate due to the
existence of the zero-modes. This is again very different from the
case of a homogeneous condensate.

The above results can be used to calculate the quantum depletion
of the soliton condensate\cite{dzi}. Let $|N_0,N_1,N_2,N_{exc}>$
denote the state with $N_0$ atoms in the condensate.  $N_1$ and
$N_2$ are number of atoms occupying the first and the second
zero-modes, and $N_{exc}$ are the number of atoms being in the
continuous modes. Assume that the initial ($t=0$) state of the
system is a Bogoliubov (quasiparticle) vacuum $|N_0,0,0,0>$. We
consider  the time evolution of the particle numbers occupying in
different modes. Noting that the particle number density operator
of the system is given by $\hat{n}(z,t)=l^{-1}
\hat{\Psi}^{\dag}(z,t)\hat{\Psi}(z,t)$ and using the expression of
the field operator,  we get the particle-number density for the
condensate mode $n_0(z)$=$<\hat{n}_0(z)>$= $l^{-1} (N_0/2)^2 {\rm
sech}^2 z$, which gives rise naturally the total particle number
$N_0$ in the condensate. Here $<\cdots>$ denotes the average over
the Bogoliubov vacuum. It is easy to show that the particle-number
densities in the first and the second zero-modes are given by
\be
\label{pnd} n_j(z,\tau)=\fr{N_0}{2l} \left[
v_j^2(z)+[u_j(z)+(-1)^j v_j(z)]^2 \tau^2\right],\;(j=1,2).
\ee
The atoms in these zero-modes are {\it incoherent} ones,
representing the quantum depletion of the soliton condensate.
There are two features for the quantum depletion: (i) The spatial
distributions of the incoherent atoms in the zero-modes are
localized because $u_j(z)$ and $v_j(z)$ are localized functions of
$z$; (ii) The distribution densities of the incoherent atoms are
time-dependent and proportional to $\tau^2$, which implies that
the depleted atoms grow {\it algebraically} as time increases and
hence {\it the soliton condensate loses atoms spontaneously even
in the absence of thermal cloud}. This is a manifestation of the
quantum diffusion of the soliton condensate implied in
(\ref{fieldoperator}). The exact expressions of the total particle
numbers depleted in the first and the second zero mode are
$N_1(t)$= $\int_{-\infty}^{\infty} dz n_1(z,t)$
=$(12+\pi)/72+2\tau^2$
and
$N_2(t)$=$\int_{-\infty}^{\infty} dz n_2(z,t)$ =$(2/3)(1+\tau^2)$.
The particle-number density in all continuous modes is given by
\be n_{exc}(z,t)=\fr{N_0}{2l} \sum_{k\neq 0} |v_k(z)|^2
=\fr{N_0}{2l} \left(\fr{1}{4}-\fr{1}{2\pi}\right){\rm sech}^4 z.
\ee
Thus, the total particle number depleted in the continuous modes
is $N_{exc}(t)$=$N_{exc}(0)$= $\int_{-\infty}^{\infty} dz
n_{exc}(z,t)$=$(\pi-2)/(3\pi)$, a small number in comparison with
that depleted in the zero modes.

We can also obtain the exact expressions of the average values for
$\hat{P}_j$ and $\hat{Q}_j$. They are given by
$<\hat{P}_1(\tau)\hat{P}_1(\tau)>$=$1/2$,
$<\hat{P}_2(\tau)\hat{P}_2(\tau)>$=$(1/2)(1+4\tau^2)$,
$<\hat{Q}_2(\tau)\hat{Q}_2(\tau)>$=$(1/2)(1+4\tau^2)$ and $
<\hat{Q}_2(\tau)\hat{Q}_2(\tau)>$=$1/2$. These expressions reflect
again the diffusion property of the soliton condensate. It should
be noted that the instability of the soliton condensate shown
above is based on a linear Bogoliubov approach. The inclusion of
the nonlinear effect between the quasiparticle excitations may
modify these results.


Our third objective is to study the stability of the soliton
condensate under a long wavelength transverse perturbation.
Assuming the confinement in the $x$ and $y$ directions is relaxed
the operator $\pa^2/\pa z^2$ in (\ref{gham}) and (\ref{solH2}) is
replaced by $\pa^2/\pa x^2+\pa^2/\pa y^2+\pa^2/\pa z^2$. To
diagonalize the Hamiltonian in this case we use the canonical
transformation $\sum_{{\bf k}_{\perp}}\sum_q \left[ u_{q,{\bf
k}_{\perp}}(z)\exp(i{\bf k}_{\perp}\cdot {\bf
r}_{\perp})\hat{c}_{q,{\bf k}_{\perp}} + v^{*}_{q,{\bf
k}_{\perp}}(z)\exp(-i{\bf k}_{\perp}\cdot {\bf
r}_{\perp})\hat{c}^{\dag}_{q,{\bf k}_{\perp}}\right]$, where ${\bf
r}_{\perp}=(x,y)$ and ${\bf q}=(k_x,k_y,q)$=$({\bf k}_{\perp},q)$
with $q=k$ (for continuous modes) or $j$ (for discrete modes). The
eigen equations (\ref{eigeneq}) are replaced by
\bea
& & \label{eigeneq1}
(\hat{L}_2-k_{\perp}^2)(\hat{L}_1-k_{\perp}^2)\phi_{\bf
q}(z)=\sigma_{\bf q} \phi_{\bf q}(z),\\
& & \label{eigeneq2}
(\hat{L}_1-k_{\perp}^2)(\hat{L}_2-k_{\perp}^2)\psi_{\bf q}(z)=
\sigma_{\bf q}\psi_{\bf q}(z),
\eea
where $\sigma_{\bf q}\equiv E_{\bf q}^2$. We consider a long
wavelength perturbation in the transverse ($x$ and $y$)
directions, thus $k_{\perp}^2$ is a small quantity which can be
taken as a small parameter to solve Eqs. (\ref{eigeneq1}) and
(\ref{eigeneq2}) using a perturbation theory. Taking $\phi_{\bf
q}=\phi_{\bf q}^{(0)}+k_{\perp}^2\phi_{\bf
q}^{(1)}+k_{\perp}^4\phi_{\bf q}^{(2)}+\cdots$, $\psi_{\bf
q}=\psi_{\bf q}^{(0)}+k_{\perp}^2\psi_{\bf
q}^{(1)}+k_{\perp}^4\psi_{\bf q}^{(2)}+\cdots$ and $\sigma_{\bf
q}=\sigma_{\bf q}^{(0)}+k_{\perp}^2 \sigma_{\bf
q}^{(1)}+k_{\perp}^4 \sigma_{\bf q}^{(2)}+\cdots$, we get a
hierarchy of the equations on $\phi_{\bf q}^{(l)}$, $\psi_{\bf
q}^{(l)}$ and $\sigma_{\bf q}^{(l)}$ ($l=0,1,2,...$). The
leading-order solutions, including the zero-modes and the
continuous modes, are just those obtained given above. We are
interested in what will happen for the zero-modes
($\sigma^{(0)}_{\bf q}=0$) when the transverse perturbation is
applied to the system. Thus in the following we consider the
dynamics of one zero-mode (say $\phi^{(0)}_{\bf
q}(z)=\phi_1(z)=u_1(z)-v_1(z)={\rm sech}z$)\cite{note1}. Because
the leading-order solutions are a set of complete and
orthonormalized functions, {\it all high-order solutions can be
obtained exactly}. In the orders $l=1$ and 2,  for frequency
correction we get $\sigma^{(1)}_{\bf
q}$=$-2\int_{-\infty}^{\infty}\phi_1^2(z)=-4$ and
$\sigma^{(2)}_{\bf q}$=$1+\int_{-\infty}^{\infty}dk a_k^{(1)}
\int_{-\infty}^{\infty}dz \left[ (k^2+1) \psi_1(z)\psi_k(z)-2
\phi_1(z)\phi_k(z)\right]=3/2$. Here
$a_k^{(1)}=-\sqrt{\pi/2}k/[(k^2+1)^2 \sinh (\pi k/2)]$. Thus under
the long wavelength transverse perturbation the eigenfrequency
corresponding to the zero-mode reads
\be \label{E1}
E_{1, {\bf k}_{\perp}}=\sqrt{-4k_{\perp}^2+3k_{\perp}^4/2}
=i2k_{\perp}\sqrt{1-3k_{\perp}^2/8},
\ee
which is pure imaginary for $k_{\perp}^2<8/3$ and hence the time
growth rate of the zero-mode is Im$(E_{1, {\bf
k}_{\perp}})$=$2k_{\perp}\sqrt{ 1-3k_{\perp}^2/8 }$. Another
zero-mode  displays also similar behavior\cite{note}.
Consequently, if there is no confinement in $x$ and $y$
directions, the soliton condensate will be dynamically unstable
under the long wavelength (i. e. for small $k_{\perp}$) transverse
perturbation. The origin of this instability comes from the
zero-modes. Thus to repress this instability a confinement in the
$x$ and $y$ directions is necessary. This is why for observing a
stable soliton condensate in experiment one must use a quasi 1D
trapping potential.

We now briefly discuss the experimental aspects of quasi 1D
condensate. Low dimensional BECs have been realized
experimentally\cite{gor} and the dark solitons in the BECs have
been investigated intensively\cite{bur,dens,hua}. Bright solitons
and bright soliton trains for attractive atomic interaction have
been observed recently\cite{kha,str}. These works will further
stimulate intensive studies on elementary excitations in
inhomogeneous Bose systems. Experimentally, a quasi-1D BEC, as
those suitable for the present work, can be obtained by tightening
the confinement in two transverse directions so that the
energy-level spacings in those directions exceed the interaction
energy. Consequently, the motion of the atoms in the transverse
directions is frozen out and the system is essentially one
dimensional. Note that the results presented above can be applied
to a quasi-1D Bose gas with attractive atomic interaction. By
assuming the 3D field operator with the form $\hat{\Phi}({\bf
r},t)$=$\Phi_0({\bf r}_{\perp})\hat{\Psi}(z,t)$, where
$\Phi_0({\bf r}_{\perp})$ is the single-particle ground state wave
function in the transverse directions, one can obtain a quasi-1D
Hamiltonian with the same form of (\ref{gham}) but the coordinate
$z$ should be scaled by $l=\hbar^2/(m I_0 |U_0^{\prime}|)$ with
$U_0^{\prime}=U_0 \int d{{\bf r}_{\perp}} |\Phi_0({\bf
r}_{\perp})|^4$.

In conclusion, we have presented rigorous results on the
diagonalization of a Bogoliubov quasiparticle Hamiltonian for a
inhomogeneous soliton condensate. Based on the set of complete and
orthonomalized eigenfunction of the Bogoliubov de Gennes
equations, we have calculated exactly the quantum depletion of the
condensate.  We have shown explicitly that two degenerate
zero-modes, appearing due to the $U(1)$ guage- and a
translational-symmetry breaking of the system, are the origin of
the quantum diffusion and the transverse instability of the
soliton condensate. We should point out that, in our study
presented above, a negligible axial trapping potential has been
assumed thus the results are valid only for a very long
cigar-shaped condensate. If the axial trapping potential, which
breaks also the translational symmetry of the system, can not be
neglected and taken as a slowly-varying function, a rigorous
result on the diagonalization of the Bogoliubov Hamiltonian and
quantum depletion of a BEC remains to be a challenge.


G. H. thanks Prof. L. You for useful discussions and the financial
support by NIST. This work was supported in part NSF-China under
Grant Nos. 90403008 and 10434060, and by the grants from Hong Kong
Research Grants Council (RGC).



\begin{references}
\bibitem{note}A detailed discussion on the completeness and the
             orthonormalities of
             the eigen-function set $\{u_q(z),v_q(z); q=k,j\}$
             and their application will be given elsewhere.
\bibitem{note1}From (\ref{eigfun1})-(\ref{zeromode2}) one has
$\phi_1(z)={\rm sech}z$, $\phi_2(z)=z\,{\rm sech}z$,
$\psi_1(z)=(1-z\tanh z)\,{\rm sech}z$,
$\psi_2(z)=\tanh z\,{\rm sech}z$,
$\phi_k(z)= e^{ikz}(1 -k^2-2ik \tanh z)/(\sqrt{2\pi}(k^2+1))$ and
$\psi_k(z)=-e^{ikz}(1+k^2+2ik\tanh z-2\tanh^2z)/(\sqrt{2\pi}(k^2+1))$.
\end{references}
\end{document}